# Comment on "Resistance of a digital voltmeter: teaching creative thinking through an inquiry-based lab"

Zoltan Gingl and Robert Mingesz

Department of Technical Informatics, University of Szeged, Árpád tér 2, 6720 Szeged, Hungary

E-mail: gingl@inf.u-szeged.hu

## Abstract
Teaching about measurements and showing examples of instruments' non-ideal behaviour are important in physics education. Digital multimeter input impedance analysis is perfect for this, since these instruments are easily available, precise enough, have good value/price ratio and represent modern technology. A recent paper (2018 *Phys. Educ.* **53** 053005) demonstrates how the investigation of the input impedance of a digital multimeter can be used in education. However, we think that some more discussion is needed to make it more complete.

## Methods to measure the input impedance

Two methods are presented in the paper [1] to measure the input impedance of a digital multimeter (DMM) in voltage measurement mode:

- Use another DMM in resistance measurement mode. In this case a constant current is forced through the input impedance and the voltage drop is measured.
- Apply a voltage source $V_S$, measure the voltage $V_{DMM}$ on a known value resistor $R$ and measure the current $I$ flowing in the circuit (see figure 1). Simple calculation gives the result:

$$Z_{IN} = \frac{R \cdot \frac{V_{DMM}}{I}}{R - \frac{V_{DMM}}{I}} \quad (1)$$

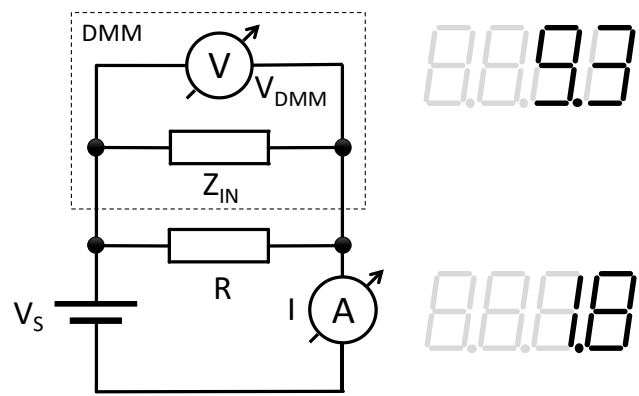

**Figure 1.** Arrangement to measure the input impedance $Z_{IN}$ of the voltmeter [1]. The readings are $V_{DMM}$=9.3 V, $I$=1.8 µA and $R$=10.6 MΩ.

The authors' observations could suggest that the accuracy is almost the same for the two methods (they obtained 10.0 MΩ and 10.1 MΩ), however, it is not the case.

## Accuracy analysis

The accuracy of the first method can be obtained using the specifications of the DMM in Ohmmeter mode. It is 2%+5digits for the DMM the authors used (BK Precision 388B) when reading is 10.00 MΩ, so the error can be as high as ±0.25 MΩ (10.00 MΩ·0.02 and 0.01 MΩ per digit). This means that the input impedance falls in the range of 9.75 MΩ to 10.25 MΩ.

The second method is much less accurate. Only a very lucky situation is shown in the paper [1], when the error of the calculated input impedance is only about 1%. Even for a perfect DMM, just due to the finite resolution, any voltage between 9.25 V and 9.35 V results the reading of 9.3 and 1.8 corresponds to the range of 1.75 µA to 1.85 µA. Figure 2 shows a simple simulation how the calculated value can vary due to this quantisation error. The voltage $V_S$ is swept in a range, the voltage $V_{DMM}$ and current $I$ are calculated and rounded to 0.1 units, then the input impedance is calculated using equation (1).

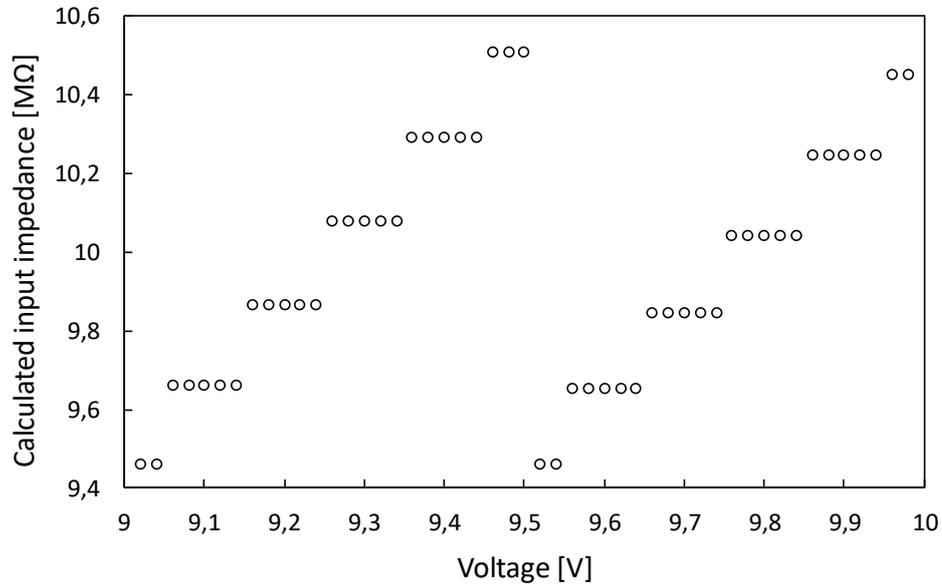

**Figure 2.** Input impedance as a function of the source voltage in a simulated measurement using $Z_{IN}$=10 MΩ and $R$=10.6 MΩ.

In addition, according to the DMM specifications, the reading of 1.8 means that the current falls in the range of 1.68 µA to 1.92 µA (error: 1%+1 digit ≈ 0.018 µA+0.1 µA≈0.12 µA). Similarly, the reading of 9.3 means that the voltage is between 8.79 V and 9.81 V (bench DMM, 0.1%+5digit ≈ 0.0093 V+0.5 V≈0.51 V). Therefore, the 9.3 V and 1.8 µA values only mean, that the input impedance is in the following range:

$$Z_{IN,min} = \frac{10.6 \cdot \frac{8.79}{1.92}}{10.6 - \frac{8.79}{1.92}} \text{MΩ} \approx 8.06 \text{ MΩ}$$

$$Z_{IN,max} = \frac{10.6 \cdot \frac{9.81}{1.68}}{10.6 - \frac{9.81}{1.68}} \text{MΩ} \approx 13.00 \text{ MΩ}$$

(2)

Assuming, that the 10.6 MΩ resistor is measured with an error of 2%+5 digits (10.338 MΩ < R < 10.862 MΩ), the range is a bit wider:

$$Z_{IN,min} = \frac{10.862 \cdot \frac{8.79}{1.92}}{10.862 - \frac{8.79}{1.92}} \text{ MΩ} \approx 7.91 \text{ MΩ}$$

$$Z_{IN,max} = \frac{10.338 \cdot \frac{9.81}{1.68}}{10.338 - \frac{9.81}{1.68}} \text{ MΩ} \approx 13.42 \text{ MΩ}$$

(3)

Since the current is low, it can't be measured accurately, this is the main source of the error. However, it is a good chance to teach about how to use the error margins specified in the manual of the DMM and also to demonstrate the effect of the quantisation error. The students can vary the voltage to see the impact on the calculated input impedance value. They can learn that it is normal to have measurement errors and that the reading is not the real value.

## An additional method

There is a third, frequently used method. A voltage source can be connected to the voltmeter via a known value resistor, see figure 3.

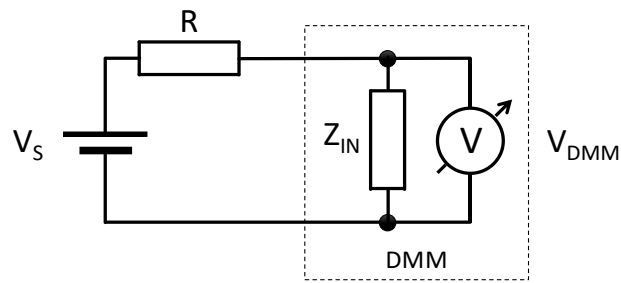

**Figure 3.** A voltage source and a resistor can be used to measure the $Z_{IN}$ input impedance.

The input impedance can be calculated as:

$$Z_{IN} = \frac{1}{\frac{V_S}{V_{DMM}} - 1} R$$

(4)

The procedure of obtaining $Z_{IN}$ can be as follows:

- apply a $V_S$ that falls in the upper half of the measurement range to have good resolution,
- measure $V_S$ directly with the DMM,
- insert a 10 MΩ 1% tolerance resistor R as shown in figure 3,
- read $V_{DMM}$ without changing the range.

For example, assuming 3.012 V and 1.507 V readings in the 4 V range and using the specifications in the datasheet (0.5%+1 digit error in voltmeter mode) we get

- 2.996 V ≤ $V_S$ ≤ 3.028 V,
- 1.498 V ≤ $V_{DMM}$ ≤ 1.516 V,
- 9.9 MΩ ≤ R ≤ 10.1 MΩ.

Substituting these into equation (4) $Z_{IN}$ can be anywhere in the range

$$Z_{IN,min} = \frac{1}{\frac{3028}{1.498} - 1} \cdot 9.9 \text{ M}\Omega \approx 9.69 \text{ M}\Omega$$

$$Z_{IN,max} = \frac{1}{\frac{2996}{1.516} - 1} \cdot 10.1 \text{ M}\Omega \approx 10.34 \text{ M}\Omega \tag{5}$$

This arrangement has some advantages:

- Only a single instrument is needed, the DMM is used in its highest accuracy mode.
- The accuracy is fairly good.
- It can be easily implemented on devices that have voltage inputs like Arduino [2,3].
- The series resistor's value can be gradually reduced to see how the voltage measurement error caused by the input impedance decreases and gets insignificant below a threshold.

## Acknowledgments

This study was funded by the Content Pedagogy Research Program of the Hungarian Academy of Sciences.